\pdfoutput=1 
\documentclass[useAMS,usenatbib]{mn2e}
\usepackage{amssymb}
\usepackage{amsmath}
\usepackage{times}
\usepackage{color}
\usepackage{graphicx}
\usepackage[T1]{fontenc}
\usepackage{aecompl}

\title[Stellar Populations of Brightest Cluster Galaxies]
  {The accretion histories of brightest cluster galaxies from their stellar population gradients}
\author[P. Oliva-Altamirano et al.]
  {Paola~Oliva-Altamirano$^{1,2}$\thanks{E-mail:poliva@astro.swin.edu.au},
  Sarah~Brough$^2$, Jimmy$^3$, Kim-Vy~Tran$^3$, Warrick~J.~Couch$^{1,2}$
  \newauthor Richard~M.~McDermid$^{2,4}$, Chris~Lidman$^2$, Anja~von~der~Linden$^{5,6,7}$,
 Rob~Sharp$^8$ \\
$^1$Centre for Astrophysics \& Supercomputing, Swinburne University of Technology, Hawthorn, VIC 3122, Australia\\ 
$^2$Australian Astronomical Observatory, PO Box 915, North Ryde, NSW 1670, Australia\\
 $^3$George P. and Cynthia W. Mitchell Institute for Fundamental Physics and Astronomy, Department of Physics and Astronomy, Texas A\&M University, \\ College Station, TX 77843, USA\\
$^4$Department of Physics \& Astronomy, Macquarie University, Sydney, NSW 2109, Australia\\
$^5$Dark Cosmology Centre, Niels Bohr Institute, University of Copenhagen, Juliane Maries Vej 30, 2100 Copenhagen ¯, Denmark\\
$^6$Kavli Institute for Particle Astrophysics and Cosmology, Stanford University, 452 Lomita Mall, Stanford, CA 94305-4085, USA\\
$^7$Department of Physics, Stanford University, 382 Via Pueblo Mall, Stanford, CA 94305-4060, USA\\ 
$^8$Research School of Astronomy \& Astrophysics, Australian National University, Cotter Road, Weston Creek, ACT 2611, Australia}
\date{Released 2002 Xxxxx XX}

\pagerange{\pageref{firstpage}--\pageref{lastpage}} \pubyear{2002}

\def\LaTeX{L\kern-.36em\raise.3ex\hbox{a}\kern-.15em
    T\kern-.1667em\lower.7ex\hbox{E}\kern-.125emX}

\begin{document}

\label{firstpage}

\maketitle

\begin{abstract}
We analyse the spatially-resolved stellar populations of 9 local ($z<0.1$) Brightest Cluster Galaxies (BCGs) observed with VIMOS in IFU mode. Our sample is composed of 7 slow-rotating and 2 fast-rotating BCGs. We do not find a connection between stellar kinematics and stellar populations in this small sample. The BCGs have shallow metallicity gradients (median $\Delta$[Fe/H]~$= -0.11\pm0.1$), high central metallicities (median $[$Fe/H$]_{[\alpha/Fe]=0}= 0.13\pm0.07$), and a wide range of central ages (from 5 to 15~Gyr). We propose that the reason for this is diverse evolutionary paths in BCGs. 67 per cent of the sample (6/9) show $\sim 7$~Gyr old central ages, which reflects an active accretion history, and 33 per cent of the sample (3/9) have central ages older than 11 Gyr, which suggest no star formation since $z=2$. The BCGs show similar central stellar populations and stellar population gradients to early-type galaxies of similar mass (M$_{dyn}> 10^{11.3}$M$_{\odot}$) from the ATLAS$^{3D}$ survey (median [Z/H]~$= 0.04\pm0.07$, $\Delta$[Z/H]~$= -0.19\pm0.1$).  However, massive early-type galaxies from ATLAS$^{3D}$ have consistently old ages (median Age~$=12.0\pm3.8$~Gyr). We also analyse the close massive companion galaxies of two of the BCGs. These galaxies have similar stellar populations to their respective BCGs.
 \end{abstract}



\section{Introduction}\label{sec:intro}
Brightest Cluster Galaxies (BCGs) are extremely luminous galaxies that are usually located in the centre of rich galaxy clusters. They have been shown to be distinct from other similarly luminous cluster galaxies \citep[e.g.][]{HAUSMAN78, POSTMAN95, LAZZATI98, LINDEN07}. In the hierarchical scenario of structure formation  \citep{TOOMRE77,WHITE78} galaxies grow in mass and size by merging with their neighbours. BCGs are predicted to have a more active merger history than lower mass galaxies \citep [][]{WHITE78, KHOCHFAR03,DELUCIA07,OSER10,NAAB13}. These galaxies are often considered as the extreme end-point of massive galaxy evolution. However, despite being among the most luminous galaxies, and generally easy to detect, observations and theory have not reached a common point yet, and their evolution is still not fully understood.  
\\\\
Observations suggest that the mass growth of BCGs evolves with time. BCGs accrete their mass at a fast rate until $z\sim0.5$, thereafter their mass growth slows down \citep{LIDMAN13, LIN13,OLIVA14,INAGAKI15}. Studies looking at BCG companions have concluded that their stellar mass grows through major mergers ($\ge1:3$ mass ratios) by a factor of $1.8\pm 0.43$ at $0.8 < z < 1.5$  \citep{BURKE13}, and mostly by minor mergers ($\leq1:4$ mass ratios) by a factor of $1.1$ at $z < 0.3$ \citep{EDWARDS12}. Major mergers are rare at low redshifts, yet still possible \citep[e.g.][]{BROUGH11,JIMMY13}.
\\\\
The recent accretion history of galaxies can be read through their stellar population gradients. In the canonical scenario, a galaxy's initial metallicity gradient is set by an initial starburst at $z\geq3$ and the metallicity decreases in the outskirts, as metallicity follows the changes in the gravitational potential \citep{SCOTT09, MCDERMID12}. This gradient can be disrupted by violent merging events (major mergers), or reinforced by minor mergers \citep{KOBAYASHI04, SPOLAOR09, COOPER10,PIPINO10}.  \citet{HIRSCHMANN14} analysed the stellar populations of 10 massive halos ($10^{12}<$~M$_{\rm halo}<10^{13}$~M$_{\odot}$) from the high-resolution cosmological  simulation of \citet{HIRSCHMANN13}. They found that major mergers do flatten the metallicity gradients. If, as predicted, BCGs have an active merger history, including several major mergers, they would be expected to have shallower metallicity gradients than lower mass galaxies. However, long-slit observations to date suggest that they have a wide range of gradients \citep{BROUGH07,LOUBSER12}. 
\\\\
Integral Field Unit (IFU) spectroscopy is a valuable tool to explore the spatially-resolved kinematics and stellar populations of galaxies. The SAURON \citep{ZEEUW02} and ATLAS$^{3D}$ \citep{CAPPELLARI11a} surveys have used IFU spectroscopy to explore a significant sample of early-type galaxies in the local Universe. \citet{KUNTSCHNER10} and \citet{ATLAS3D} presented the stellar population analysis of the SAURON and ATLAS$^{3D}$ samples, respectively, finding that 40 per cent of the galaxies typically show contributions from young stellar populations connected to low mass, fast rotator systems. In contrast, they find that slow rotators are generally consistent with old ($\geq 10$ Gyr) stellar populations. The most massive systems (stellar mass $\geq 10^{10.5}$M$_{\odot}$) have the flattest metallicity gradients. However, the ATLAS$^{3D}$ sample contains 21 galaxies with dynamical masses greater than $10^{11.3}$M$_{\odot}$, and only one of those is a BCG (M87).
\\\\
\citet{JIMMY13} analysed the kinematics and photometry of  a sample of 10 BCGs observed with the VIMOS IFU, of which 4 have close massive companions. If BCGs were the product of many minor mergers they would be expected to be slow-rotating galaxies. \citet{JIMMY13} found that 30 per cent of the BCGs in their sample are fast rotators. The simulations of \citet{NAAB13}, predict that angular momentum mostly depends on the gas content of the galaxies involved in the interaction, suggesting that the slow or fast rotation could be a temporary state. \citet{JIMMY13}, also find through photometric analysis \citep[$G-M_{20}$;][]{LOTZ08} that 40 per cent of the galaxies in the sample have undergone a minor merger within the last 0.2 Gyr. 
\\\\
Our paper is the third in a series analysing the spatially-resolved spectra of BCGs \citep[after][]{BROUGH11,JIMMY13}. We present here the spatially$-$resolved stellar populations of the BCG sample presented in \citet{JIMMY13}. This is the first IFU analysis dedicated to the stellar populations of BCGs. We will investigate whether BCG accretion histories are different from those of the general massive early-type galaxy population by comparing our measurements to those from SAURON and ALTLAS$^{3D}$.
\\\\
We present the sample selection and observations in Section 2. Section 3 describes the stellar kinematic and photometric measurements from \citet{JIMMY13}. It also describes the early-type galaxy samples of SAURON \citep{KUNTSCHNER10} and ATLAS$^{3D}$ \citep{ATLAS3D} to which we will compare our sample BCGs throughout the paper. Section 4 presents the stellar population analysis. Section 5 summarises the main results. These are later discussed in Section 6. Our conclusions are presented in Section 7. The cosmology adopted throughout this paper is H$_0$=70 kms$^{-1}$ Mpc$^{-1}$, $\Omega_M$=0.3, $\Omega_{\Lambda}$=0.7.

\section{Observations}
In this section we summarise the observations made and the data reduction process. These are described in full in \citet{JIMMY13}. 
 \subsection{Spectroscopic Measurements}\label{sec:obs}
 The BCG sample is selected from \citet{LINDEN07}. The galaxies are part of the C4 cluster catalogue (Miller 2005) of the third data release of the Sloan Digital Sky Survey \citep[SDSS; ][]{YORK00}. These observations consist of 10 BCGs, 4 of which have massive companions within $\sim 10^{\prime\prime}$ (corresponding to $\sim 18$~kpc at $z<0.1$). We use the same nomenclature as \citet{JIMMY13}, i.e. we present each cluster as the last 4 digits in the SDSS flag, rather than SDSS-C4-DR3 \textit{number}. The galaxies were observed with the Very Large Telescope using the IFU mode of the VIMOS spectrograph \citep[][]{LEFEVRE03}, with the high-resolution blue grism, which has a spectral resolution of 0.51~\AA/pixel. The observations were made in two sets, April to August of 2008 and April to July of 2011 (Prog.~ID 381.B-0728 and Prog.~ID 087.B-0366, respectively). The galaxies used in this study have a spatial sampling of $0.67^{\prime\prime}$/pixel with a field-of-view (FOV) of  $27^{\prime\prime} \times 27^{\prime\prime}$. The average seeing of the observations is $0.9^{\prime\prime}$. The rest wavelength range is $\sim3900$~to~$5600$~\AA.  
\subsection{IFU Data Reduction}\label{sec:redu}
The IFU data reduction consists of two different stages: (a) The VIMOS Pipeline \citep{IZZO04} which generates the calibrations files (fibre identification, master bias, etc.), and does a first order flux calibration. This flux calibration corrects the spectrum shape, using the standard stars observed on the same night as the galaxies. The VIMOS FOV is formed by four quadrants. The VIMOS pipeline reduces each quadrant separately. As a result we obtain the science spectrum for each spaxel in each quadrant. (b) We use our own IDL routines to mask the bad fibres, subtract the sky for each quadrant and then combine the quadrants into a three dimensional data cube. The multiple exposures for each observation are combined using a 5$\sigma$ clipped median. This code is publicly available\footnote{http://galaxies.physics.tamu.edu/index.php/Jimmy\#Code}. Finally, we flux calibrate the spectrum using the photometric standard stars. 

\section{Previous measurements}
\citet{JIMMY13} measured the stellar kinematics and photometry of the galaxies in our sample, we summarise their method and results in this section.
\subsection{Stellar Kinematics}\label{sec:ppxf}
First a signal-to-noise ratio (SNR) cut of 5 across all the spaxels (spatial pixels) was applied. The spaxels were then re-binned to a minimum SNR of 10, using the spatial binning Voronoi code of \citet{CAPPELLARI03}. The velocity and line-of-sight velocity dispersion were computed using the penalised fitting scheme of \citet[pPXF;][]{PPXF}, and the MILES \citep[Medium-resolution Isaac Newton Telescope Library of Empirical Spectra; ][]{MILES} library stellar templates. pPXF fits the stellar library templates to the absorption line features of the BCG spectra, giving the redshifts and the broadening of the spectral lines.
\\\\ 
\textbf{The angular momentum} was characterised by the $\lambda_{\rm R}$ parameter defined by \citet{EMSELLEM07}. It is calculated as follows:    
\begin{equation}
\lambda_{\rm R} \sim \frac{\langle R |V| \rangle}{\langle R\sqrt{V^2+\sigma^2 \rangle}},
\end{equation}
where $R$ represents the radius of the galaxy, V is the stellar velocity and $\sigma$ the velocity dispersion. The numerator and denominator are luminosity weighted. A higher $\lambda_{\rm R}$ represents a higher angular momentum. \textbf{The ellipticity ($\epsilon$)} at the effective radius of each galaxy was measured using the publicly available IDL routine $find\_galaxy.pro$ developed
 by Michele Cappellari\footnote{http://www-astro.physics.ox.ac.uk/$\sim$mxc/idl/}. Following \citet{EMSELLEM11}, the values of $\lambda_{\rm R}$ and $\epsilon$ can be used to distinguish fast and slow rotators (FR and SR respectively) by using the threshold: 
\begin{equation} \label{eq:FR}
\lambda_{\rm R} \geq (0.31 \pm 0.01)\times \sqrt{\epsilon},
\end{equation}  
where FRs lie above this threshold and SRs lie below.\\\\
\textbf{The dynamical mass} was measured using the standard equation given in \citet{CAPPELLARI06}.  
\begin{equation}\label{eq:mass}
M_{dyn}=\frac{5R_e \sigma_e^2}{G},
\end{equation}
where $\sigma_e$ is the aperture corrected velocity dispersion of the integrated spectrum within the effective radius; G is the gravitational constant. In Table \ref{tab:kin} we summarise the relevant kinematic results. 7 of the BCGs are SR and 2 are FR. For $\sigma_{e}, \epsilon_e$ and $\lambda_{R_e}$ we refer the reader to Table 2 of \citet{JIMMY13}.
\subsection{Photometric Analysis}
\citet{JIMMY13} analysed the photometry of this sample using images from SDSS Data Release 3. They measured the effective radius by fitting a 2D de Vaucouleurs profile. They also analysed the presence of recent mergers using the Gini, and M$_{\rm 20}$ coefficients \citep{LOTZ08}. This method studies the distribution of light looking for irregularities that could indicate morphological signatures of mergers. A galaxy is a merger candidate if it crosses the threshold:
\begin{equation}\label{eq:merger}
G \geq -0.14M_{20} + 0.33
\end{equation}
Where M$_{\rm 20}$ is the 2nd order moment of the brightest 20~per~cent of pixels, and G is the Gini coefficient.  In the case of gas-poor galaxies like our sample, a galaxy will be above the threshold if it is currently merging or has merged in the last 0.2~Gyr \citep{LOTZ11}. 4 of the BCGs in the sample are merging. The relevant kinematic and photometric results of \citet{JIMMY13} are presented in Table \ref{tab:kin}.
\begin{table}
\begin{minipage}{0.85\linewidth}\label{sec:phot}
\caption{Kinematic properties of BCGs and their companions from \citet{JIMMY13}. Seven of the BCGs are slow rotating (SR) and two are fast rotating (FR). Four of the BCGs show photometric signs of merging.}

\label{tab:kin}
\begin{tabular}{@{}lccccc}
\hline
Galaxy&$z$&log M$_{\rm dyn}$~M$_{\odot}$&R$_e(^{\prime\prime})$&\small{Merging?}&FR/SR\\
& &\tiny{$\pm0.01$}&\tiny{$\pm 0.01$} &\tiny{$G-M_{20}$}&\\
\hline\hline
BCGs&&&&&\\[0.1cm]
1027A& 0.090 &11.79& 6.98 & y &SR\\
1042& 0.094 &11.83& 7.22 & n & SR\\
1050 & 0.072 &11.78& 8.43 & n & SR\\
1066& 0.083 &11.62& 5.07 & y & SR\\
2001& 0.041 &11.38& 5.84 & n & SR\\
2039& 0.082 &11.86& 8.82 & n & SR\\
2086& 0.083 &11.60& 4.83 & y & SR\\
1048A& 0.077 &11.59&5.17&y&FR\\
1261& 0.037 & 11.32& 5.76 & n & FR\\
\hline
Comp&&&&&\\[0.1cm]
1027B& 0.090 & 11.17& 4.39 & y &FR\\
1048B& 0.080 &10.51&1.08&y&FR\\
1048C& 0.074 &10.54&1.24&y&FR\\
\hline
\medskip
\end{tabular}\\
\end{minipage}
\end{table}

\subsection{Stellar populations from the SAURON and ATLAS$^{3D}$ samples}\label{sec:sauron}
Throughout the paper we compare our observations to those of early-type galaxies of similar mass observed by ATLAS$^{3D}$ (which includes the SAURON galaxy sample). The ATLAS$^{3D}$ sample is composed of 260 field and cluster early-type galaxies and it only contains one BCG, M87. The spatially-resolved stellar populations (central values and gradients) of the SAURON sample were presented in \citet{KUNTSCHNER10}. The central stellar populations of the ATLAS$^{3D}$ sample were presented in \citet{ATLAS3D}. We therefore compare our central stellar populations with the whole ATLAS$^{3D}$ sample and the stellar population gradients only with the SAURON sample.
\\\\
\citet{KUNTSCHNER10} and \citet{ATLAS3D} use the stellar models of \citet{SCHIAVON07} in the Lick/IDS system \citep{WORTHEY97} to measure the stellar population parameters of age, total metallicity [Z/H] and abundance of alpha elements [$\alpha$/Fe]. For our spectral fitting, we use models with only solar abundances, i.e.  $[\alpha$/Fe$]=0$. In this case [Fe/H] is effectively a measure of the total metallicity\footnote{[Fe/H] = [Z/H] - 0.75*[$\alpha$/Fe] \citep{CONROY12} }. Therefore, we directly compare the ATLAS$^{3D}$ total metallicities to our measured metallicities throughout the paper. The central stellar populations in the ATLAS$^{3D}$ sample correspond to an aperture of 0.125~R$_e$.
\\\\
We compare the median, and standard deviation for the two samples in the same mass range (M$_{dyn}>10^{11.3}$M$_{\odot}$). In order to compare our BCGs with a non-BCG early-type galaxy sample, we do not include the BCG M87. This gives a comparison sample of 20 massive early-type galaxies from the ATLAS$^{3D}$ sample. We highlight M87 as a blue-filled circle in the figures.
\section{Stellar Population Analysis}\label{sec:annuli}
\begin{figure*}
\includegraphics[width=0.7\linewidth]{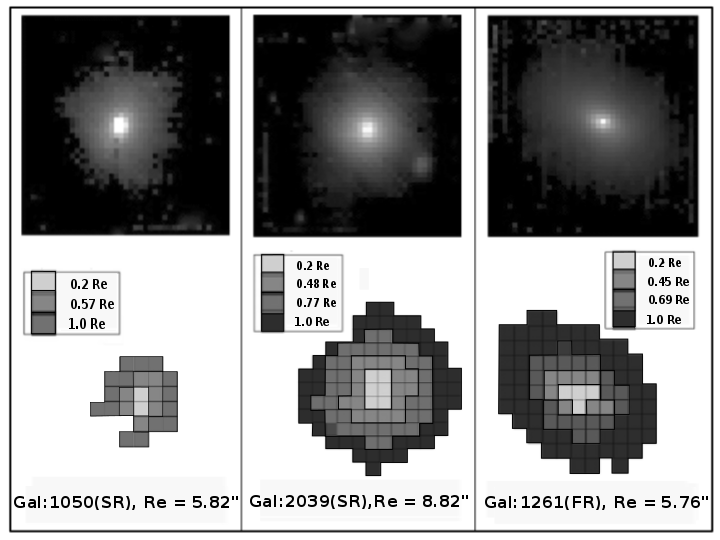}
\caption{Three representative BCGs: 1050 (SR), 2039 (SR), 1261 (FR). \textbf{Upper panels}: The flux-collapsed VIMOS image ($27^{\prime \prime}\times27^{\prime \prime}$).  \textbf{Lower panels}: The annular distribution for each galaxy. The central annulus, corresponds to the aperture $0.2\pm03$~R$_e$. The annuli all extend to 1~R$_e$.}  
\label{fig:FOV}
\end{figure*}

The study of stellar populations requires high SNR spectra. In order to secure a high enough SNR we use our own python routine\footnote{http://astronomy.swin.edu.au/$\sim$poliva/codes/annuli\_stacking/spectra.py}  to stack the spaxels within annuli for each galaxy. To identify the annuli we follow the total flux in the wavelength$-$integrated galaxy image. This indirectly maps the galaxy morphology, i.e. the shape of the annuli is determined by the galaxy's morphology. Each spectrum is shifted to rest frame wavelength before stacking using the velocity measurements obtained from pPXF. The spectra are then broadened to a reference velocity  dispersion, $\sigma$ (the maximum velocity dispersion of the galaxy). This reduces the dilution of the spectral features due to rotational broadening and allows us to have a fixed and constant velocity dispersion when measuring the stellar population parameters. As a result, we have one spectrum per annulus per galaxy.  
\\\\
From the 10 BCGs and 4 companion galaxies presented in \citet{JIMMY13}, the BCG 1153, and the companion of BCG 1066 have too low SNR for stellar population analysis. Our final sample thereby consists of 9 BCGs, 4 of them with close massive companions. For 2 of these (1027, 1048) it was possible to resolve the companions as well. In those cases we refer to the main galaxy as 1027A and 1048A, and the companions as 1027B, 1048B, and 1048C. 
\\\\
In Fig \ref{fig:FOV} we show the annular distribution of 3 representative galaxies. The upper panels are the flux-collapsed VIMOS image. The lower panels are the annular distribution per galaxy. The annuli cover up to 1 R$_e$ in each galaxy. The central aperture has been defined as $0.2\pm0.03$~R$_e$, allowing the central annulus to contain two or more spaxels (for most of the galaxies). The median value of the SNR of the final sample stacked spectra is $\sim35$~\AA$^{-1}$ with the majority having SNR~$> 20$~\AA$^{-1}$. Only the outermost annuli in the companion galaxies 1048B and 1048C drop below this threshold (SNR~$= 15$~\AA$^{-1}$). 
\\\\
The analysed rest wavelength range (3900-5600~\AA) covers the absorption line indices: Ca$_{\rm K}$ (Ca II 3933), Ca$_{\rm H}$ (Ca II 3968), H$\delta$, H$\gamma$, Ca$_{\rm G}$ (Ca I 4307), H$\beta$, Fe5015, Mgb5175, Fe5270. Our age and metallicity measurements are based on the full spectrum. However, we mask the weaker Balmer indices H$\delta$, and H$\gamma$ to ensure a clean comparison with the ATLAS$^{3D}$ data (which age measurements are based on H$\beta$). The enhanced lines Ca$_{\rm K}$, Ca$_{\rm H}$, and Ca$_{\rm G}$ provide a strong case for the existence of old stellar populations. Furthermore, where there are mixed stellar populations, the ratio between these lines indicates the relative importance of the young stellar populations \citep{SANCHEZ12}.
\\\\
In the following subsections we describe the method we implement to estimate the metallicities and ages.
\subsection{Stellar Population Models}\label{sec:miles}
We use the stellar population models of \citet{VAZDEKIS10}, based on the MILES stellar library \citep{MILES},  to estimate the ages and metallicities of galaxies in our sample.  The \citet{VAZDEKIS10} models use the Padova 2000 \citep[][]{GIRARDI00} isochrones which cover a metallicity range of  [Fe/H]~\footnote{[Fe/H] = log$_{10}$ (Z/Z$_{\odot}$), and Z$_{\odot} = 0.0189$ \citep{ANDERS89}} = [-1.7, 0.4] and ages [$2.1\times 10^7,1.7\times 10^{10}$] yr. After testing our spectra with the whole age range, we restrict our analysis to single stellar population sequences of [$1\times 10^9,1.7\times10^{10}$] yr, divided into 40 age bins. 
\\\\
We also considered the stellar population models of PEGASE-HR \citep{LEBORGNE04} coupled to the ELODIE \citep{ELODIE} stellar library. These models are known for their high spectral resolution (0.55~\AA) and for allowing the user to choose the initial mass function along with other physical ingredients.  Its wavelength coverage is consistent with our data, ($\lambda>3900$~\AA). Furthermore, the stellar library has the same age range as MILES. We find both libraries to give similar results. Both metallicities and ages are consistent within 1$\sigma$ uncertainties. However, 
the flux calibration of the MILES library over a wide spectral range is more appropriate for our spectral fitting of unresolved stellar populations. The average spectral resolution of MILES ($2.3$~\AA) is also closer to that of the VIMOS observations ($2.1$~\AA). This minimises information loss when broadening the library spectra. We therefore use the \citet{VAZDEKIS10} models in the results presented here. 
\subsection{Full Spectrum Fitting}\label{sec:steckmap}
\begin{figure}
\includegraphics[width=\linewidth]{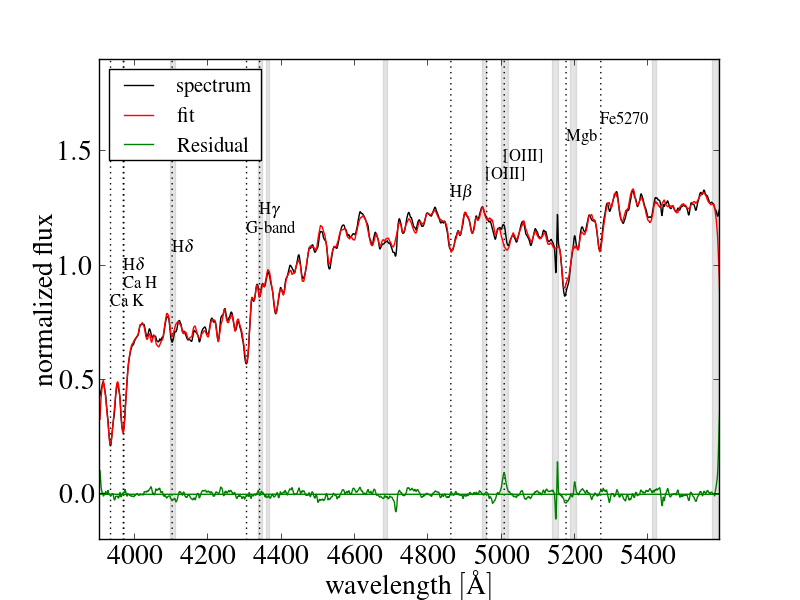}
\caption{Rest wavelength spectrum of the central annulus of BCG 2039. The black line represents our data, the red line shows the best fitting stellar populations from STECKMAP. The green line shows the residuals of the fit. The shaded regions show the regions that were masked on the fit. The major features are marked with dotted vertical lines. The recovered stellar populations are $[$Fe/H$]_{[\alpha/Fe]=0}=0.13\pm0.01$ and Age~$=9.6\pm1.0$~Gyr. The fit is a good match to the data.}   
\label{fig:spectrum}
\end{figure}

To estimate the stellar population parameters we used the full spectrum fitting technique \citep[e.g.][]{KOLEVA08}. This technique features some advantages over classical methods, i.e. the Lick/IDS system \citep[e.g.][]{WORTHEY97}, as it exploits all the information contained in the spectra, pixel by pixel, independently of the spectrum shape. It also allows analysis at medium spectral resolutions ($<3$~\AA), in contrast to the Lick/IDS system which has a low spectral resolution, ($>8$~\AA). We use the STEllar Content and Kinematics via Maximum A Posteriori algorithm \citep[STECKMAP;][]{STECKMAP,STECMAP} to extract the ages and metallicities from our spectra.
\\\\
STECKMAP uses Bayesian statistics to estimate the stellar content of the spectra. It is based on a non-parametric formalism. The code returns a luminosity-weighted age and metallicity distribution. This comes from flux-normalising the stellar library (rather than mass normalisation). The results are a proxy to the star formation history of the galaxy. The method is regularised by a Laplacian kernel in order to avoid chaotic oscillations. To prevent systematic errors from poor flux calibration, the code produces a non-parametric transmission curve which represents the instrumental response multiplied by the interstellar extinction. We further mask the spectral region around the weak emission lines so that they do not interfere with the stellar population model fitting (e.g, [NeIII] 3868.71, [OIII] 4959, 5007, [NI] 5198, 5200).
\\\\
In order to obtain accurate, robust results we first test the spectrum by measuring the stellar population parameters using two different age initial conditions: (a) a flat stellar age distribution. (b) a random Gaussian distribution of ages \citep[see][]{OCVIRK11}. We then compare the results from both runs expecting them to be consistent. If this test is successful, we proceed to calculate the final value of the stellar population parameters.  The final estimated luminosity-weighted age and metallicity are the median values from 150 Monte Carlo realisations. In each Monte Carlo realisation the initial condition is a random Gaussian age distribution that is later refined through iterations until it reaches the best fit. The measurement uncertainties are the standard deviations of the 150 Monte Carlo realisations.   In Fig \ref{fig:spectrum} we show the spectrum of the central annulus of 2039. The black line is our data, the red line represents the best fit to the data, the vertical dotted lines indicate the significant lines. 
\\\\
Unfortunately, STECKMAP, in conjunction with the MILES/ELODIE stellar libraries, has the disadvantage of being tied to solar abundance ratios ($[\alpha$/Fe$]=0$). This means that were not include very high metallicities and $\alpha$-element [$\alpha$/Fe] abundance in this model. BCGs are known to have high metallicities and super solar [$\alpha$/Fe] ratios \citep[e.g.][]{LINDEN07}. We therefore explore the impact of this on our fits using a Lick index analysis.

\subsection{Lick/IDS System}
\begin{figure}
\includegraphics[width=1\linewidth]{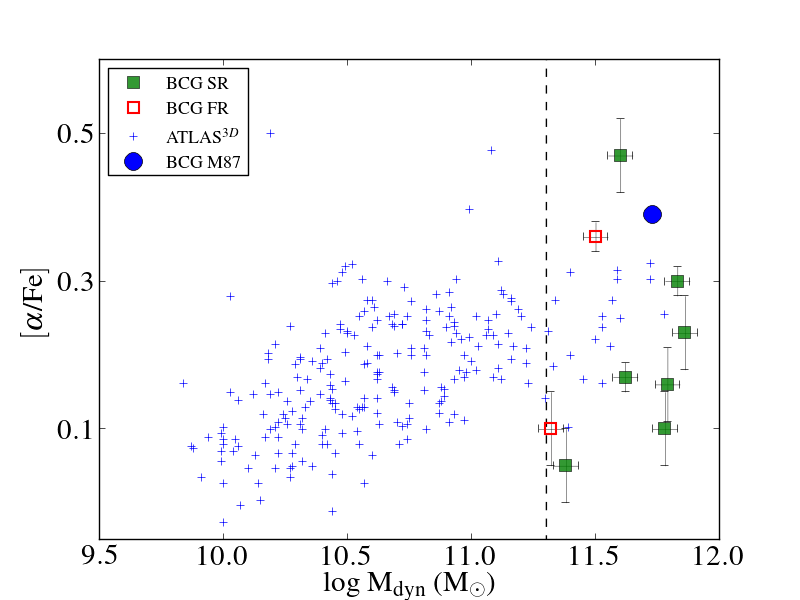}
\caption{Central $\alpha$-abundance $[\alpha$/Fe$]$ as a function of galaxy dynamical mass. The blue crosses represent the early-type galaxies from the ATLAS$^{3D}$ \citep{ATLAS3D} survey with M87 shown as a blue-filled circle. The slow rotator BCGs are shown as green-filled squares. The fast rotator BCGs are shown as red-open squares. The dashed line shows the mass range (M$_{dyn}> 10^{11.3}$M$_{\odot}$) used in the comparison between the two samples. We find that the BCGs have similar $\alpha$-enhanced ratios (median [$\alpha$/Fe]~$=0.17\pm0.04$) to the early-type galaxies (median [$\alpha$/Fe]~$=0.24\pm0.03$).}  
\label{fig:alpha}
\end{figure}

\begin{figure}
\begin{minipage}[b]{0.95\linewidth}
\includegraphics[width=1\linewidth]{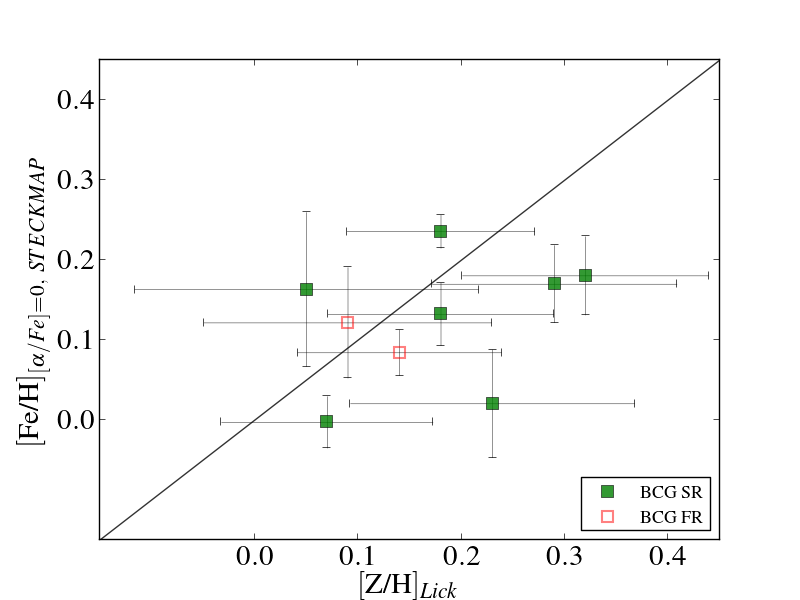}
\end{minipage}
\begin{minipage}[b]{0.95\linewidth}
\includegraphics[width=1\linewidth]{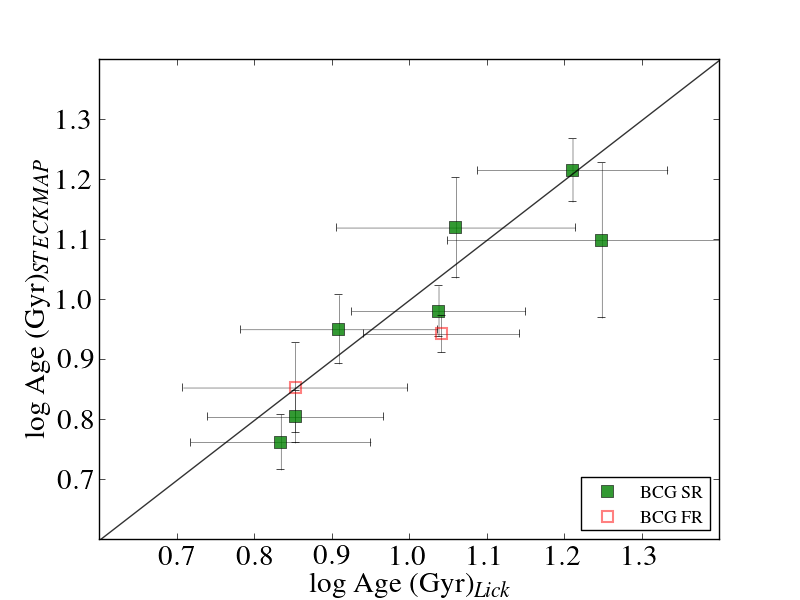}
\end{minipage}
\caption{Comparison between STECKMAP and Lick estimated stellar populations. The green-filled squares represent the central stellar populations of the SR BCGs and the red-open squares represent the FR BCGs. The line shows the one-to-one relationship. The $[$Fe/H$]_{[\alpha/Fe]=0}$ is equivalent to [Z/H] (see section 3.3). Our STECKMAP and Lick measurements agree within 1$\sigma$ error. }   
\label{fig:met_age}
\end{figure}

The stellar population parameters can also be estimated using the Lick/IDS system  \citep[e.g.][]{WORTHEY97}. This method measures the equivalent width of the absorption features at a fixed IDS resolution ($>9$~\AA) to later compare them with stellar population models that provide the corresponding age and metallicity values \citep[e.g.][]{WORTHEY97,VAZDEKIS97, PROCTOR02, SCHIAVON07}. 
\\\\
To probe the robustness of our results obtained using STECKMAP we test our VIMOS spectra with the Lick/IDS system (see also Appendix A). We use the same stellar models and absorption lines as the ATLAS$^{3D}$ team to measure the [$\alpha$/Fe] abundances, ages and metallicities i.e. \citet[][]{SCHIAVON07}, H$\beta$, Fe5010, and Mgb. In Fig \ref{fig:alpha} we show the central [$\alpha$/Fe] of our BCGs (green-filled and red-open squares) and the ATLAS$^{3D}$ sample (blue crosses) as a function of mass. We find the BCG [$\alpha$/Fe] values (median [$\alpha$/Fe]~$=0.17\pm0.04$) to be consistent with those of the ATLAS$^{3D}$ massive galaxies (median [$\alpha$/Fe]~$=0.24\pm0.03$).  
\\\\
In Fig \ref{fig:met_age} we show the estimated metallicities and ages from both methods, STECKMAP and Lick indices. The Lick ages (median Age~$=10.0\pm1.1$~Gyr) and metallicities (median [Z/H]~$=0.18\pm0.02$) are consistent with those from STECKMAP (median Age~$=8.9\pm3.3$~Gyr, median $[$Fe/H$]_{[\alpha/Fe]=0}=0.13\pm0.07$) within 1$\sigma$ error. Our STECKMAP results are also consistent with the stellar populations analysed by \citet{GALLAZZI05} from the SDSS DR4 spectra (of the galaxies in our sample).
\\\\
We adopt as the final result the stellar population parameters measured with STECKMAP. We discuss these results in the following sections.
\section{Results}
We present here the main results from our BCG stellar population analysis. 
\subsection{Central Stellar Populations}\label{sec:age_met}   
\begin{figure*}
\includegraphics[width=0.6\linewidth]{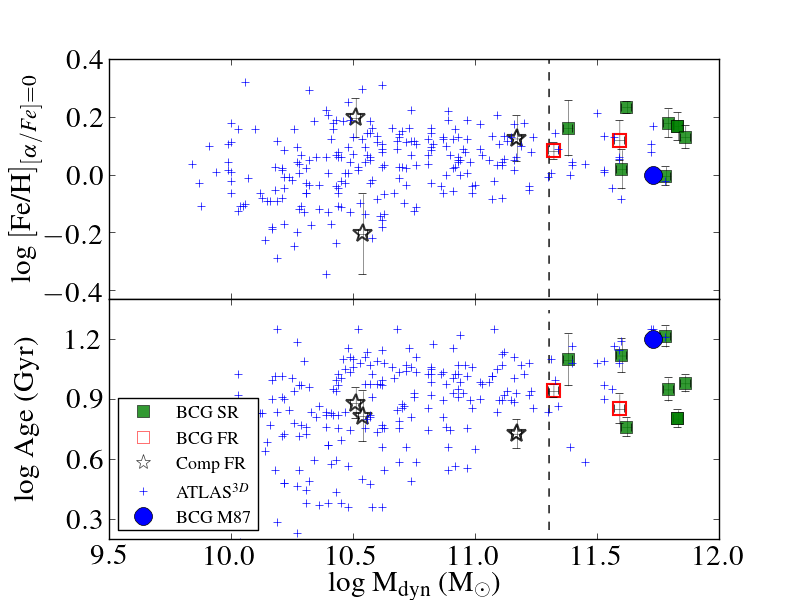}
\caption{Central stellar populations as a function of dynamical galaxy mass. The SR BCGs are shown as green-filled squares. The FR BCGs are shown as red-open squares. The FR companion galaxies are shown as open stars. The blue crosses represent the central stellar populations of the early-type galaxies in the ATLAS$^{3D}$ sample. The blue-filled circle is the BCG M87 from the ATLAS$^{3D}$ sample. The dashed line shows the mass range (M$_{dyn}> 10^{11.3}$M$_{\odot}$) used in the comparison between the two samples. \textbf{Upper panel:} Central metallicities as a function of dynamical galaxy mass. The BCGs show similar metallicities (median $[$Fe/H$]_{[\alpha/Fe]=0}= 0.13\pm0.07$) compared to the ATLAS$^{3D}$ galaxies at the same mass (median [Z/H~]$= 0.04\pm0.07$). \textbf{Lower panel:} Central ages as a function of mass. The BCGs central ages (median Age~$=8.9\pm3.3$) are consistent with the ages of the ATLAS$^{3D}$ galaxies of the same mass (median Age~$=12.2\pm1.3$) within $2\sigma$ error. The central stellar populations of BCGs show little scatter or dependence on their angular momentum.}
\label{fig:central}
\end{figure*}

Our stellar population measurements come from a luminosity-weighed distribution, therefore our results are sensitive to the brightest and youngest population of stars in the galaxy \citep[e.g.][]{TRAGER09}. The central values are measured in an aperture of $0.2\pm0.03~$R$_e$ and are given in Table \ref{tab:central}. We find that 6 out of 9 BCGs (67 per cent of the sample) have central intermediate ages (5 Gyr < Age < 10 Gyr), and 3 out of 9 (33 per cent of the sample) are centrally old (Age > 11 Gyr). The median central age is $8.9\pm3.3$~Gyr. The BCGs have homogeneous super-solar metallicities (median $[$Fe/H$]_{[\alpha/Fe]=0}= 0.13\pm0.07$) in their central regions.
\\\\
Fig \ref{fig:central} shows the central metallicities (upper panel) and ages (lower panel) as a function of galaxy mass. The SR and FR BCGs are presented as green-filled squares and red-open squares  respectively. The companion galaxies (all FRs) are shown as open stars. We do not find any significant difference between the SRs and FRs, consistent with \citet{ATLAS3D}. The crosses represent the central stellar populations of the early-type galaxies in the ATLAS$^{3D}$ sample \citep[][]{ATLAS3D}. From a Kolmogorov~-~Smirnov (K-S) test we find that the hypothesis that the massive early-type galaxy metallicities and BCG metallicities come from the same distribution, can not be rejected at a 10 per cent confidence level (P-value 0.11). The BCGs have similar central metallicities (median $[$Fe/H$]_{[\alpha/Fe]=0}= 0.13\pm0.07$) to early-type galaxies in ATLAS$^{3D}$ (median [Z/H]~$= 0.04\pm0.07$) at fixed mass. 
\\\\
The BCGs have slightly younger central ages (median Age~$=8.9\pm3.3$~Gyr) compared to the ATLAS$^{3D}$ galaxies at the same mass (median Age~$=12.0\pm3.8$~Gyr). From a K-S test we find that the age distribution of early-type galaxies and BCGs are different (P-value $1.2\times10^{-6}$). We also find that the BCGs have similar central stellar populations to the companion galaxies.
\\\\
In summary, our sample BCGs show homogeneous central metallicities and a wide range of ages. They show similar metallicities to other early-type galaxies of similar mass.
\begin{table}
\begin{minipage}{0.85\linewidth}
\caption{Central stellar populations of BCGs and their companions.}
\label{tab:central}
\begin{tabular}{@{}llcccc}
\hline
&Galaxy&log [Fe/H]&log [Fe/H]&Age&\small{Age error}\\
&&&error&(Gyr)& (Gyr)\\
\hline\hline
BCG SR&1027A&0.18&0.049&8.9&1.1\\
&1042&0.17&0.049&6.4&1.1\\
&1050&-0.003&0.032&16.4&1.1\\
&1066&0.235&0.021&5.8&1.1\\
&2001&0.163&0.097&12.6&1.3\\
&2039&0.132&0.039&9.6&1.1\\
&2086&0.02&0.068&13.2&1.2\\
BCG FR&1048A&0.121&0.069&7.1&1.2\\
&1261&0.084&0.029&8.8&1.1\\
\hline
BCG&median&0.132&0.048&8.9&1.1\\
\hline\hline
Comp&1027B&0.127&0.08&5.4&1.2\\
&1048B&-0.203&0.139&6.6&1.3\\
&1048C&0.198&0.069&7.6&1.2\\
\hline
Comp & median&0.126&0.079&6.5&1.2\\
\hline\hline
\medskip
\end{tabular}
\textbf{NOTE:} The errors listed are the measurement uncertainties. 
\\
\end{minipage}
\end{table}
\begin{table}
\begin{minipage}{0.85\linewidth}
\caption{Stellar population gradients of BCGs and their companions.}
\label{tab:gradients}
\begin{tabular}{@{}llcccc}
\hline
&Galaxy&$\Delta$ [Fe/H]& $\Delta$ [Fe/H] &$\Delta$ Age&$\Delta$ Age\\
&& &error && error\\
\hline\hline
BCG SR&1027A&-0.115&0.08&0.004&0.061\\
&1042&0.028&0.094&0.019&0.162\\
&1050&0.092&0.042&-0.197&0.066\\
&1066&-0.414&0.158&0.054&0.116\\
&2001&-0.054&0.225&-0.134&0.092\\
&2039&-0.171&0.09&0.03&0.036\\
&2086&0.05&0.026&-0.091&0.102\\
BCG FR&1048A&-0.213&0.221&0.135&0.119\\
&1261&-0.198&0.096&0.034&0.045\\
\hline
BCG& median &-0.115&0.094& 0.019& 0.092\\
\hline\hline
Comp&1027B&-0.17&0.006&0.214&0.089\\
&1048B&0.174&0.35&0.066&0.061\\
&1048C&-0.11&0.163&0.133&0.125\\
\hline
Comp & median&-0.110&0.160&0.133&0.088\\
\hline\hline
\medskip
\end{tabular}
\textbf{NOTE:} The errors listed are the measurement uncertainties. 
\end{minipage}
\end{table}
\begin{figure}
\begin{minipage}[b]{1.05\linewidth}
\includegraphics[width=1\linewidth]{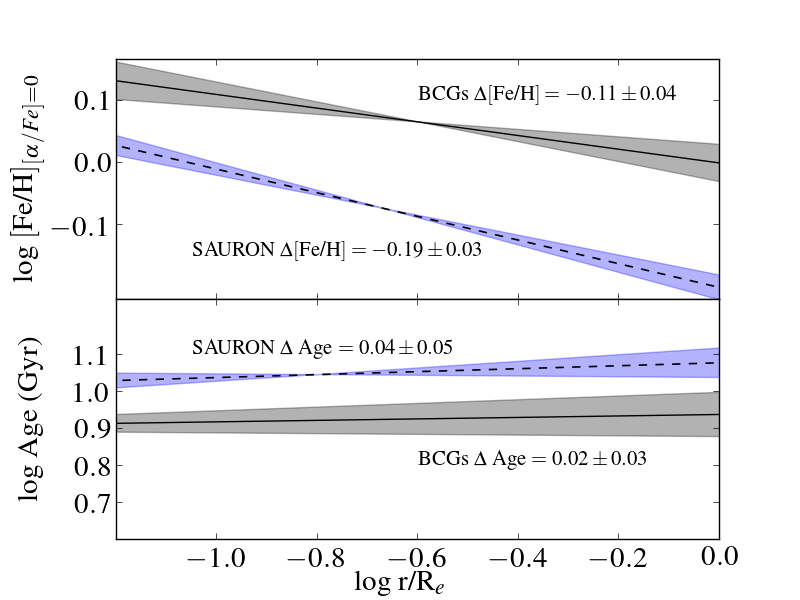}
\end{minipage}
\caption{Stellar population profiles of BCGs and early-type galaxies of similar mass. The solid lines represent the median stellar population gradients of the BCGs in our sample. The dashed lines represent the median stellar population gradients of the most massive galaxies in the SAURON sample. The grey (BCGs) and blue (SAURON) shaded regions indicate the error on the median. These values are specified in each panel. BCGs and early-type galaxies at the same mass have similar gradients.}
\label{fig:grad}
\end{figure}
\subsection{Age and Metallicity Profiles}\label{sec:profiles}
\begin{figure*}
\begin{minipage}[b]{0.6\linewidth}
\includegraphics[width=1\linewidth]{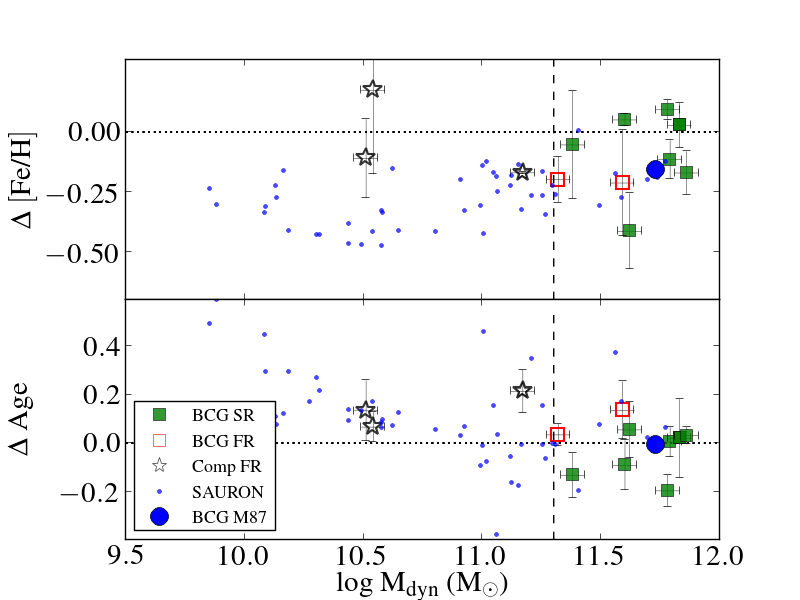}
\end{minipage}
\caption{Metallicity (upper panel) and age (lower panel) gradients as a function of dynamical galaxy mass. The SR BCGs are shown as green-solid squares. The FR BCGs are shown as red-open squares. The blue dots represent the galaxies in the SAURON sample. The blue-filled circle is the BCG M87 from the SAURON sample. The dotted line indicates a flat gradient. The dashed line shows the mass range (M$_{dyn}> 10^{11.3}$M$_{\odot}$) used in the comparison between the two samples. The BCGs ($\Delta$[Fe/H]~$=-0.11\pm0.1$) and the most massive early-type galaxies ($\Delta$[Fe/H]~$=-0.19\pm0.1$) have similar stellar population gradients.}
\label{fig:grad_mass}
\end{figure*}

The stellar population gradients and their uncertainties are measured using a linear log-log chi-squared fitting routine. The uncertainties are the standard deviation of the fit. We summarise the stellar population gradients ($\Delta$ [Fe/H] and $\Delta$ Age) in Table~\ref{tab:gradients}. The metallicity and age profiles for the BCGs and companion galaxies are illustrated in Appendix B. Most of the BCGs in our sample have shallow metallicity gradients, $\Delta$[Fe/H]~$>-0.3$, except for 1066 which has a gradient of $\Delta$[Fe/H]~$-0.41\pm0.1$. The median value is $\Delta$[Fe/H]~$=-0.11\pm0.1$. The companion galaxies also show shallow metallicity gradients. However 1048B and 1048C have R$_e$ values close to the seeing FWHM, which may act to dilute the measured gradient in these galaxies.
\\\\
We summarise the stellar population gradients we observe for the BCGs in Fig \ref{fig:grad} and compare that to the gradients of the SAURON galaxies at the same mass range. The offset between the two profiles illustrates the differences between the central stellar populations in the two samples. 
\\\\
In Fig \ref{fig:grad_mass} we show the metallicity (upper panel) and age (lower panel) gradients as a function of galaxy mass. The BCGs have shallow metallicity (median $\Delta$[Fe/H]~$= -0.11\pm0.1$) and age gradients (median $\Delta$Age~$=0.02\pm0.03$), similar to those of the SAURON early-type galaxies at the same mass ($\Delta$[Fe/H]~$=-0.19\pm0.1$, $\Delta$Age~$=0.04\pm0.05$). We do not find any correlation between the stellar kinematics and the stellar population gradients.

\section{Discussion}\label{sec:disc}
BCGs represent the extremely massive end of the early-type galaxy population. These galaxies live in high-density environments commonly surrounded by many companions. We have presented here the first integral field analysis of the radial stellar populations of 9 BCGs up to 1~R$_e$. 
\subsection{Stellar Ages}
Hydrodynamical simulations of early-type galaxies (in less dense environments than BCGs) predict more massive galaxies to be older than less massive galaxies, such that at masses $>10^{10.5}$~M$_{\odot}$ the galaxies are older than $10$~Gyr \citep[e.g.][]{NAAB13,PEEPLES14,HIRSCHMANN13} and show passive evolution from $z=2$. We find that 3 out of 9 BCGs in our sample have old central ages ($> 12$ Gyr), in agreement with this prediction. These 3 galaxies are also consistent with the massive early-type galaxies from ATLAS$^{3D}$ (median Age~$=12.0\pm3.8$~Gyr).
\\\\
However, 6 out 9 BCGs in our sample have central intermediate ages (5 Gyr $\leq$ Age < 10 Gyr). Previous observations have shown that these intermediate ages in  BCGs are not unusual. \citet{LOUBSER09} analysed a large sample of 49 BCGs, and found that 24 of them (49~per~cent of the sample) are younger than 9~Gyr old. \citet{FITZPATRICK14} found that, at fixed velocity dispersion and surface brightness, central galaxies in SDSS are younger than satellite galaxies. \citet{BARBERA14} also found that central galaxies have younger ages and higher metallicities than isolated early-type galaxies. Consistently, these 6 BCGs sit at the younger age limit of the massive early-type galaxies from ATLAS$^{3D}$ (Fig~\ref{fig:central}). 
\\\\
Many hypotheses have tried to explain why such massive galaxies as BCGs have intermediate age stellar populations. A common prediction is that gas cooling from the intra-cluster medium may be forming stars which will result in young ages \citep[e.g.][]{EDWARDS07,ODEA08, BILDFELL08, LOUBSER14}. For most of the clusters considered here, X-ray imaging is not available, so that we cannot assess whether they host a cool core or not. Only two of our galaxies have been confirmed to be hosted by a non-cooling flow cluster \citep{WHITE97} and for those we find contradictory results: 2001 is one of the oldest galaxies in our sample, consistent with the cool core hypotheses, however, 1066 shows a young central region and a positive age gradient, suggesting that cool cores are not the explanation for the intermediate ages observed. 
\\\\
The semi-analytical model of \citet{TONINI12} focusses on BCGs rather than on the general population of early-type galaxies, predicting that BCGs have more prolonged star formation as a result of their active merger history. The BCGs experience continuous bursts of star formation across cosmic time, generating many stellar populations super imposed on one another. Our age measurements are luminosity-weighted, which means they reflect the youngest stellar population of the galaxy. Therefore, the intermediate ages we find suggest that the last star formation event could have taken place at $z\sim1$ when galaxy mergers are more likely to be gas rich. 
\subsection{Metallicities} 
The BCGs in our sample show very homogenous central metallicities (median $[$Fe/H$]_{[\alpha/Fe]=0}= 0.13\pm0.07$). These high metallicities are consistent with previous long-slit and fibre observations of BCGs \citep{BROUGH07,LINDEN07,LOUBSER09,EIGENTHALER13} and are in agreement with the hypothesis of continuous star formation events at high redshifts. 
\\\\
We find that the BCGs we observe have a range of metallicity gradients, from flat to shallow (median $\Delta$[Fe/H]~$= -0.11\pm0.1$) similar to other massive early-type galaxies at similar mass ($\Delta$[Z/H]~$=-0.19\pm0.1$; Fig \ref{fig:grad_mass}). 
\\\\
Hydrodynamical simulations predict that galaxies that form through dissipative core collapse have typical metallicity gradients $\Delta$[Fe/H]~$\sim-0.4$ \citep[e.g][]{KOBAYASHI04,HIRSCHMANN14}. This is significantly steeper than the gradients we observe. However, simulations also show that this initial gradient can later be affected by accretion of external stellar populations, i.e. mergers \citep{HIRSCHMANN13,MARTIZZI14}.
\\\\
Dissipationless major mergers make the stars lose their orbits and move randomly within the distribution of the galaxy, inducing stellar population mixing and flattening the gradients \citep{HOPKINS092}. This suggest that the BCGs as well as the massive SAURON galaxies have gone through at least one recent dissipationless major merger since $z<1$ \citep{KOBAYASHI04,HIRSCHMANN14}.  
\subsection{Merger Histories}
From our analysis we conclude that BCGs have diverse evolutionary paths. 3 out of 9 BCGs in our sample show old and metal-rich central stellar populations, and shallow metallicity gradients. This suggest that their stars were formed in-situ at $z>2$. Thereafter the galaxies grow in mass and size by at least one major merger and many minor mergers \citep[e.g.][]{KOBAYASHI04,HIRSCHMANN14}. These galaxies are similar to the massive early-type galaxies in the SAURON and ATLAS$^{3D}$ sample which also have old central stellar populations and shallow metallicity gradients.  
\\\\
The rest of the sample (6 out of 9) BCGs have intermediate central ages, high central metallicities, and shallow metallicity gradients in BCGs. This implies that these galaxies have experienced active accretion histories throughout cosmic time, as predicted by semi-analytical and dark matter simulations \citep{DELUCIA07, TONINI12, LAPORTE13}. The dense environment where BCGs evolve allows them to experience many mergers. These mergers will trigger star formation at high redshifts, and will disrupt the metallicity gradients at $z<1$, given that the fraction of gas in the merging galaxies decreases with time. 
\\\\
\citet{JIMMY13} found that 4 of the 9 BCGs studied here show photometric signatures of minor mergers. The effect of minor mergers are not apparent in the inner ($<1$~R$_e$) stellar population gradients studied here, as minor mergers only affect stellar population gradients at $>2$~R$_e$ \citep[][]{FOSTER09,BARBERA12,PASTORELLO14, HIRSCHMANN14}. However, the photometric results are evidence of the active merging activity of these galaxies. Furthermore, 4 of the galaxies in our sample have close massive companions, most of these companions are FRs and are gravitationally bound to their respective BCG. This suggests a potential future major merger \citep{JIMMY13}. 
\\\\
Many studies have found similar results on the stellar populations of BCGs \citep[e.g.][]{WHILEY08, STOTT08}. \citet{WEN11} analysed the BCGs colours from different high-redshift data sets (CFHT, COSMOS, SWIRE) and found that BCGs are consistent with stellar population synthesis models in which the galaxy formed at $z>2$. However, a large fraction of the sample shows bluer colours on the g'~-~z' and B~-~m$_{\mu m}$ bands at $z\sim0.8$, indicating star formation at those epochs. Furthermore, some BCGs show low levels of star formation in the local Universe \citep[e.g.][]{LIU12,OLIVA14,FRASER14}. This is consistent with the hypothesis that BCGs have complex accretion scenarios.
\\\\
Thanks to the spatial extent of the IFU spectroscopy we were able to resolve 3 companion galaxies (1027B, 1048B,  1048C) from 2 of the BCGs (1027A, 1048A). We find that the companion galaxies similar stellar populations with their respective BCG. However, due to the fact that their effective radius are close to the seeing FWHM, their stellar population gradients are unreliable.
\subsection{Connection Between Stellar Populations and Kinematics}
One of the advantages of using IFU spectroscopy for this analysis is that we can compare the stellar populations to the kinematics of the galaxies in our sample. We have analysed 7 slow and 2 fast rotating BCGs and their kinematics appear to be independent of their stellar populations. We do not find any correlation between the angular momentum of the galaxies and their stellar population gradients in this small sample. The SRs show a large scatter in their metallicity gradients. The 2 FRs have similar metallicity gradients, but within the range of the SRs. 
\\\\
\citet{MOODY14} and \citet{NAAB13} showed that if the gas fraction in a major merger is less than $10$~per~cent (dissipationless), the galaxy tends to maintain the slow or fast rotation of their progenitors. The BCGs in our sample are likely to have preserved the slow or fast rotation of their progenitors, given the lack of gas observed at present epochs and the evidence of recent dissipationless major mergers.

\section{Conclusions}
We present the first analysis of the stellar populations of 9 BCGs using IFU observations. We compare their stellar populations to those of the IFU observations of field early-type galaxies from the SAURON and ATLAS$^{3D}$ sample. We draw the following conclusions:
\\\\
(1) The BCGs have a wide range of stellar ages, high metallicities, and have shallow metallicity gradients. This implies diverse evolutionary paths (passive and active accretion). The BCGs' central stellar populations and gradients are consistent with those of early-type galaxies of similar mass, with the exception that the BCGs have a wide range of central ages. 
\\\\
(2) Three of the BCGs have similar mass close companions galaxies within 18~kpc. From those 3 BCGs we were able to resolve the companions of 2 of them. The companion galaxies have central stellar populations consistent with their respective BCG.
\\\\
(3) We do not observe a relationship between the stellar populations of BCGs and their stellar kinematics (slow and fast rotators).
\\\\
IFU analysis has allowed us to determine the angular momentum of BCGs and to study their stellar population gradients without the orientation bias innate to long-slit spectroscopy. It has also allowed us to include their companion galaxies in the analysis. This study hints of intriguing differences between BCGs and similar mass early-type galaxies but requires much larger samples to confirm. The SAMI galaxy survey \citep{FORGARTY14, ALLEN14} currently underway will allow us to achieve this goal. New ultra wide-field IFU spectrographs such as MUSE on the Very Large Telescope and large date sets as the MASSIVE survey \citep[][]{MA14} will also allow us to determine whether the picture is different at radii beyond 1~R$_e$.

\section{ACKNOWLEDGEMENTS}
P.O.A. acknowledges the valuable input of Michaela Hirschmann, Ignacio Ferreras, Harald Kuntschner, and Marja Seidel. 
\\\\We thank our referee, Reynier Peletier, for his helpful feedback, which helped to probe the accuracy of our stellar population analysis.     
\\\\The Dark Cosmology Centre (DARK) is funded by the Danish National Research Foundation.

\bigskip
\setlength{\bibhang}{2.0em}
\setlength\labelwidth{0.0em}
\bibliographystyle{mn2e}
\bibliography{POA}
\newpage
\bigskip

\section*{Appendix A: Representative Lick indices equivalent width}
\begin{figure*}
\begin{minipage}[b]{0.4\linewidth}
\includegraphics[width=1\linewidth]{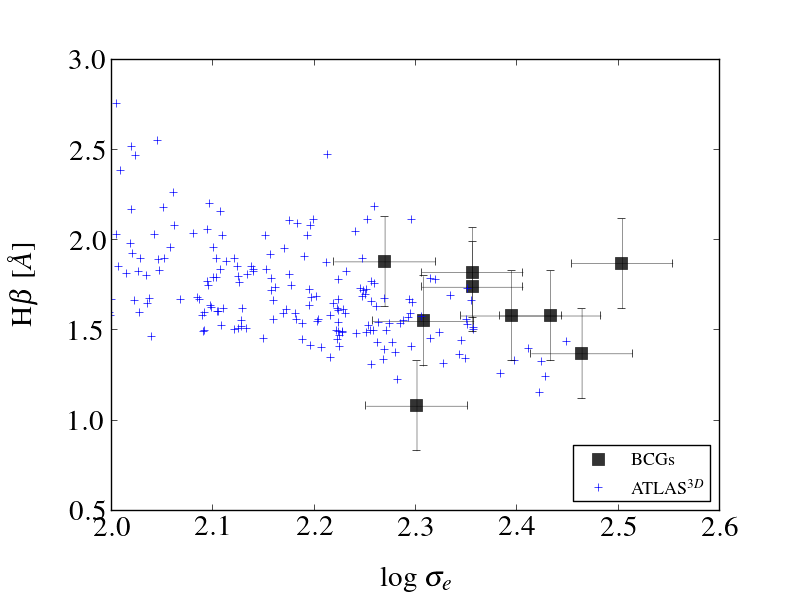}
\end{minipage}
\begin{minipage}[b]{0.4\linewidth}
\includegraphics[width=1\linewidth]{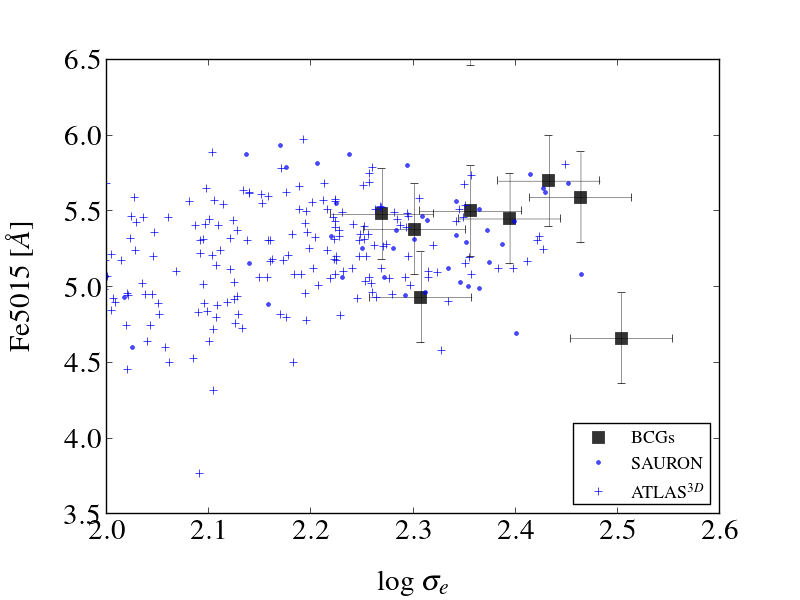}
\end{minipage}
\begin{minipage}[b]{0.4\linewidth}
\includegraphics[width=1\linewidth]{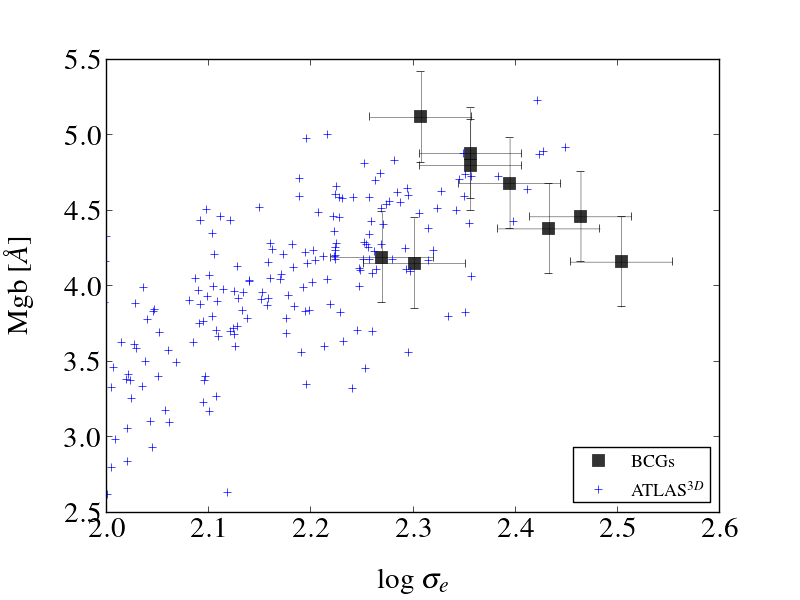}
\end{minipage}
\caption{Lick indices as a function of velocity dispersion. The squares represent the BCGs and the blue crosses represent the ATLAS$^{3D}$ galaxies. The indices were measured at a Lick/IDS resolution ($>8.4$ \AA). The BCGs show higher H$\beta$ and Fe5010, and similar Mgb values than the early-type galaxies at fixed velocity dispersion}
\label{fig:indices}
\end{figure*}

In Fig \ref{fig:indices} we show the indices H$\beta$, Fe5010, and Mgb, proxies of age and metallicity respectively, as a function of velocity dispersion. The BCGs show higher H$\beta$ and Fe5010, and similar Mgb values than the early-type galaxies at fixed velocity dispersion.
\section*{Appendix B: Stellar population profiles of brightest cluster galaxies and their companions}
\begin{figure*}
\begin{minipage}[b]{0.8\linewidth}
\includegraphics[width=1\linewidth, height=6.8cm ]{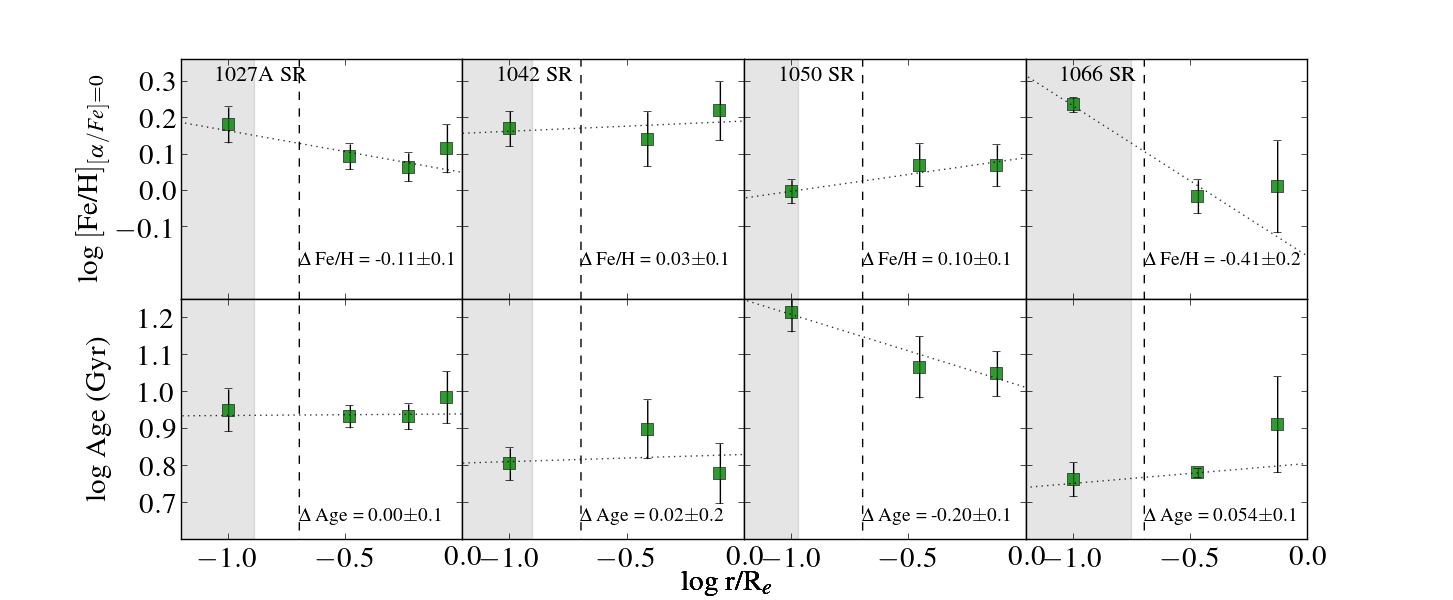}
\end{minipage}
\begin{minipage}[b]{0.65\linewidth}
\includegraphics[width=1\linewidth, height=6.8cm]{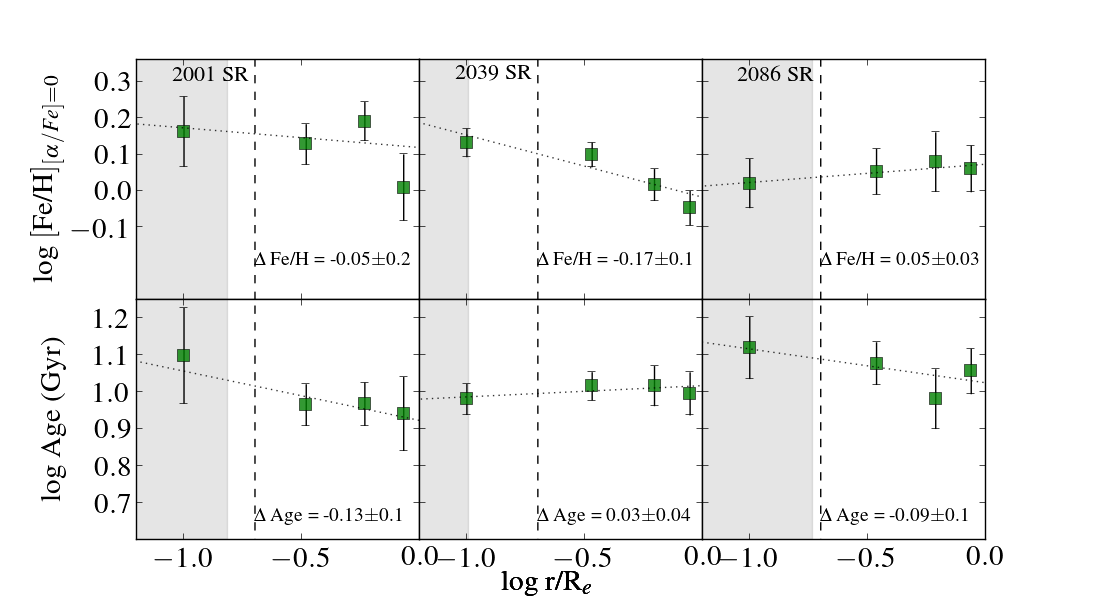}
\end{minipage}
\caption{Metallicity (upper panel) and age (lower panel) profiles of the SR BCGs. The name of the galaxy can be found in the upper-left corner. In the lower-right corner we show the gradient values. The shaded area represents the seeing FWHM ($0.9^{\prime \prime}$). Each green square represents the metallicity and age value of each annulus in the galaxy. The dotted line represents the best fit to the profile. The dashed line indicates the central region (an aperture of 0.2~R$_e$). . The SR BCGs have a large scatter in their observed metallicity and age gradients.}
\label{fig:profiles0}
\end{figure*}

\begin{figure*}
\begin{minipage}[b]{0.5\linewidth}
\includegraphics[width=1\linewidth,height=6.8cm]{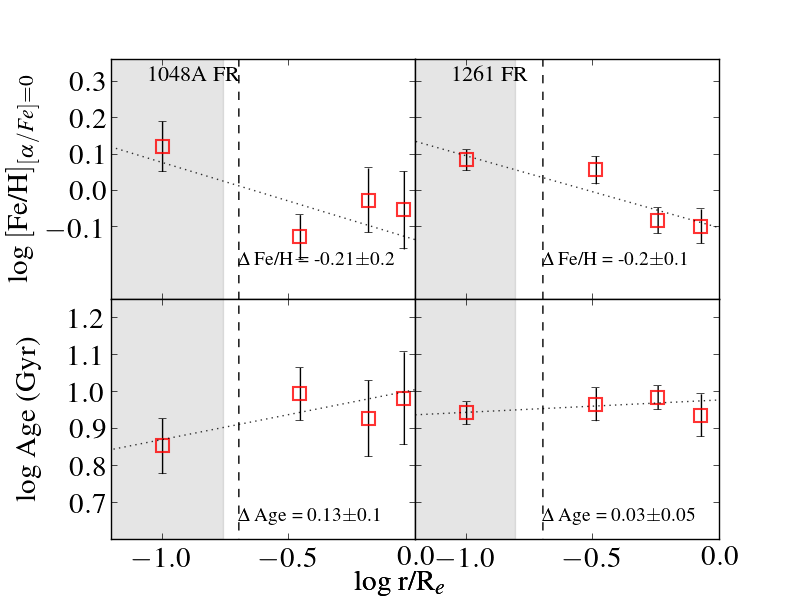}
\end{minipage}
\caption{Metallicity (upper panel) and age (lower panel) profiles of the FR BCGs. The name of the galaxy can be found in the upper-left corner. In the lower-right corner we show the gradient values. The dotted line represents the best fit to the profile. The dashed line indicates the central region (an aperture of 0.2~R$_e$). The shaded area represents the seeing FWHM ($0.9^{\prime \prime}$). Each red-open square represents the metallicity and age value of each annulus in the galaxy. Both of the FR BCGs have negative metallicity gradients and positive age gradients. }
\label{fig:profiles1}
\end{figure*}

\begin{figure*}
\begin{minipage}[b]{0.7\linewidth}
\includegraphics[width=1\linewidth]{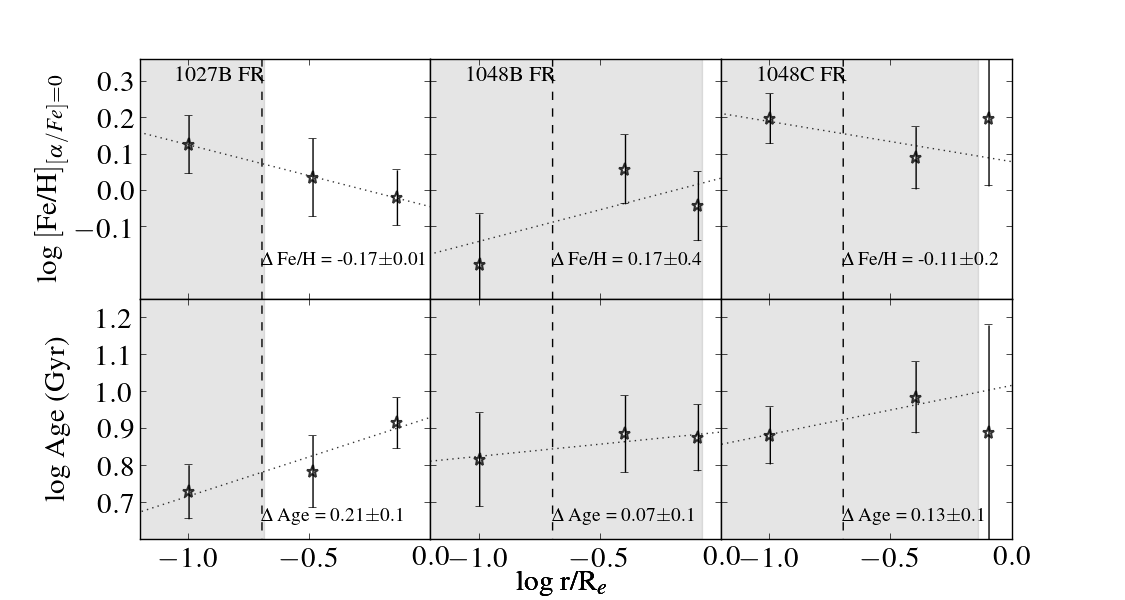}
\end{minipage}
\caption{Metallicity and age profiles of the companion galaxies (all FRs). Each double panel shows the metallicity (upper panel) and age (lower panel) profiles of each companion. The name of the galaxy can be found in the upper-left corner. The dotted line represents the best fit to the profile. The dashed line delimitates the central region (an aperture of 0.2~R$_e$). The shaded area represents the seeing FWHM ($0.9^{\prime \prime}$). Each open star represents the metallicity and age value of each annulus in the galaxy. All 3 of them have a flat metallicity gradient. However, in 1048B and 1048C, this could be a result of the seeing FWHM being equivalent to the R$_e$ in these galaxies. }
\label{fig:profiles2}
\end{figure*}

In this Section we show the stellar population profiles for each of the SR BCGs (Fig \ref{fig:profiles0}), FR BCGs (Fig \ref{fig:profiles1}), and  FR companion galaxies (Fig \ref{fig:profiles2}). 

\end{document}